\documentclass[prl,aps,twocolumn,showpacs]{revtex4-1}
\usepackage{amsmath}
\usepackage{graphicx,amssymb}
\usepackage[pdftex]{color}
\usepackage{color}
\begin{document}
\bibliographystyle{apsrev}

\title{Comment on "Correlation between Bulk Thermodynamic Measurements
and the Low-Temperature-Resistance Plateau in $\bold {SmB_6}$"}

\author{Kai Chen$^{1,*}$, Jean-Paul Kappler$^1$}

\address{\rm $^1$Synchrotron SOLEIL, L'Orme des Merisiers,Saint-Aubin-BP48, 91192 GIF-sur-YVETTE CEDEX,France}

\date{\today}

\begin{abstract}
Low-temperature-resistivity plateau observed in $\rm SmB_6$ single crystal,which is due to surface, not bulk, conduction has been confirmed from electrical transport measurements. Recently, the correlation between bulk thermodynamic measurements and the low-temperature-resistance plateau in $\rm SmB_6$ have been investigated and a change in Sm valence at the surface has been obtained from x-ray absorption spectroscopy and x-ray magnetic circular dichroism. Here we show that the statement of the report are not supported by the results from x-ray absorption spectroscopy and x-ray magnetic circular dichroism.    
\end{abstract}
\maketitle

In a recent article, W. A. Phelan and co-workers\cite{Phelan2014} report data on the correlation between bulk thermodynamic measurements and the low-temperature-resistance plateau in $\rm SmB_6$. They found surface conductivity of $\rm SmB_6$ increases systematically with bulk carbon content and addition of carbon is linked to an increase in n-type carriers, larger low-temperature electronic contributions to the specific heat and a broadening of the crossover to the insulating state. A change in Sm valence at the surface has been obtained from x-ray absorption spectroscopy(XAS) and x-ray magnetic circular dichroism(XMCD), which is claimed to be the definitive proof of changes in the electronic structure at the surface of $\rm SmB_6$. This statement is true while the data from XAS and XMCD are problematic which may misleading the further investigations. 

In their report, the XAS and XMCD of surface and bulk are obtained from total electron yield (TEY) and fluorescence yield(FY), which are believed to be sensitive to the surface and bulk of the sample, respectively. However, the "surface" is not well defined here, which should be the electron escaping lengh in the order of $\rm \sim 2nm$ while the "bulk" is related to the thickness of $\sim10$ times higher. Besides, nothing is reported for the $\rm SmB_6$ surface treatment which is important for the XAS measurement since naturally oxidized or cleaved surface are quite different. Furthermore, at the $\rm M_5$ edge of Sm, XAS from FY may be quite different from that obtained from TEY and transmission\cite{Pompa1997}, due to the 3d core hole lifetime broadening $\rm \Gamma$ dominated by the auger decay. However, such a deviation between TEY and FY is neglected in $\rm SmB_6$ shown in Fig.8a in\cite{Phelan2014}, in which the peaks from $\rm Sm^{2+}$ and $\rm Sm^{3+}$ are well distinguished. 

\begin{figure}[b]
\includegraphics[width=\linewidth]{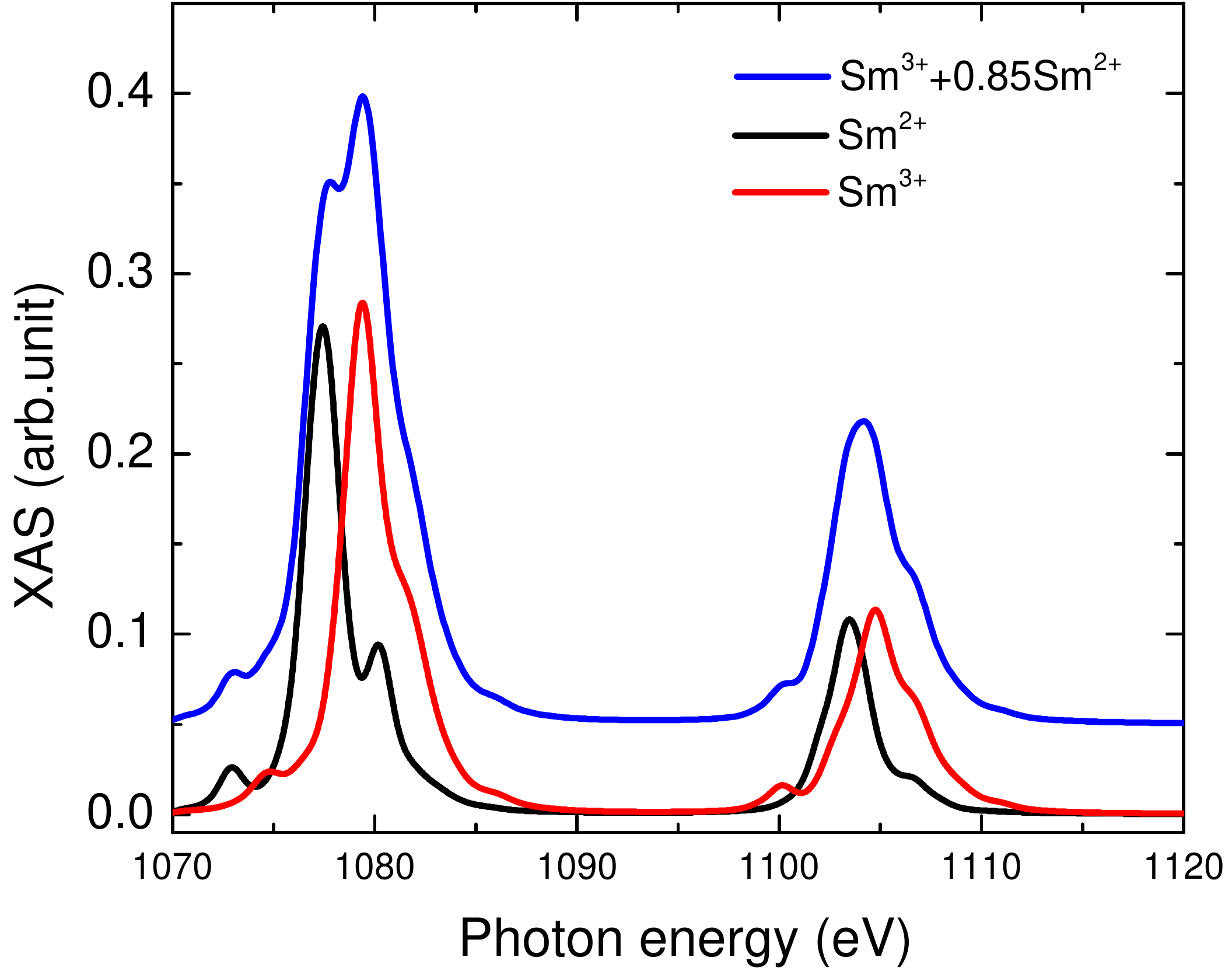}
\caption{\small{ XAS of $\rm Sm^{2+}$ and $\rm Sm^{3+}$ from atomic multiplet calculation using CTM4XAS\cite{Stavitski2010}. The electrostatic and exchange parameters were scaled down to 80\% of the atomic Hartree-Fock value. } }
 \label{fig:1}
\end{figure}

As they claimed in Fig.8 in \cite{Phelan2014}, ``the bulk spectra measured from FY (red curve) are consistent with a mixture of $\rm Sm^{2+}$ and $\rm Sm^{3+}$, with no appreciable magnetization, as previously reported\cite{Tarascon1980, Mizumaki2009}". Nothing related to magentism is reported in the ref\cite{Tarascon1980, Mizumaki2009}. It is also claimed: ``in contrast, the surface spectra from TEY (black curve) consists of almost entirely $\rm Sm^{3+}$ and shows a discernible XMCD signal characteristic of a net magnetic moment, approximately 1/10 of that observed in ferromagnetic $\rm Sm_{0.974}Gd_{0.02}Al_2$ \cite{Dhesi2010}". Here the spectra of $\rm Sm^{2+}$ and  $\rm Sm^{3+}$ are mistaken in the report. To clarify, the XAS of $\rm Sm^{2+}$ and $\rm Sm^{3+}$ obtained from atomic multiplet calculation using CTM4XAS\cite{Stavitski2010} are shown in Fig.1. The electrostatic and exchange parameters were scaled down to 80\% of the atomic Hartree-Fock value. In the case of $\rm Sm^{3+}$ ions, because the two first excited, J =7/2 and J=9/2, multiplets are relatively close in energy to the fundamental J=5/2 multiplet \cite{Kramida2014}, it is necessary to account for the crystalline electric field effects, not only on the fundamental, but also on these excited multiplets. The mixing of these higher multiplets into the fundamental leads to a reduction of the magnetic moment \cite{Buschow1973}. Such an effect is not considered in the calculation since only a slightly change of the XAS shape will be observed \cite{Dhesi2010}. 
The XAS of $\rm Sm^{2+}$ is left shifted compared to that of $\rm Sm^{3+}$ in Fig.1, which is normal and attributed to the chemical shift, and similar to the previous results from experimental results \cite{Kaindl1984} and theoretical calculation\cite{Thole1985}. However it is opposite in Fig.8 in\cite{Phelan2014}, where the surface state of $\rm Sm^{3+}$ is left shifted. This needs to be corrected at least by an erratum.

According to the experimental data of XAS, it is not possible that the surface is in pure $\rm Sm^{2+}$ state while the bulk is mixed with $\rm Sm^{2+}$ and $\rm Sm^{3+}$.  For Sm metals, which in the bulk is a trivalent of $\rm Sm^{3+}$ at the surface was turned into divalent configuration of $\rm Sm^{2+}$ \cite{Wertheim1978,Allen1978,Johansson1979}. For $\rm SmB_6$, surface valence between 2.5 and 2.6 was determined from X-ray photoemission spectroscopy(XPS)\cite{Heming2014}.  Interestingly, as we calculated the shape of XAS for the surface is more like a $\rm 4f^5$ ground state with $\rm Sm^{3+}$, not a $\rm 4f^6$ ground state with $\rm Sm^{2+}$ (Fig.1). The XMCD results are also puzzling. Indeed, a magnetic TEY signal is observed in the ``surface" case ($\rm Sm^{3+}$ like), with the same shape as $\rm Sm^{3+}$ in\cite{Dhesi2010}, whereas no magnetic signal has been detected for the bulk. As the bulk is a $\rm Sm^{2+}$-$\rm Sm^{3+}$ mixing, at least the Sm atoms in the 3+ state should give a magnetic response.

There still remains the problem with the shape of XAS from FY. The intensive peak at higher energy cannot be understood. We doubt there is the energy shift in the FY XAS since the shape canbe well fitted with the XAS of $\rm Sm^{3+}$ and $\rm Sm^{2+}$, as shown in Fig.1. In this case, the XAS and XMCD data canbe well understood and supports their statements very well. However, we have no idea if there exists the energy shift between the XAS from TEY and FY shown in Fig.8a in\cite{Phelan2014} and needs to be checked by the authors.  

We conclude that the XAS and XMCD spectra of Sm in\cite{Phelan2014} are problematic. Several possible mistakes have been considered to understand the results, among which the energy shift between the XAS from TEY and FY may be the explanation. We also doubt if chemical states of Sm canbe determined from the comparison from the XAS of $\rm M_{4,5}$ edge measured from TEY and FY\cite{Pompa1997}.

\end{document}